\documentclass[12pt]{article}

\usepackage{fancyhdr}            
\fancypagestyle{plain}{%
  \fancyhead{}                                     
  \fancyfoot[CE,CO]{\thepage}       
  \fancyhead[CE,CO]{\textcolor[rgb]{0.64,0.15,0.15}{Published in \emph{Journal of Elasticity} \textbf{126}, 215--229
  (2017)
\\
doi: http://dx.doi.org/10.1007/s10659-016-9590-5}}
\setlength{\headheight}{14.5pt}}

\usepackage{latexsym}
\usepackage{amsmath,amsfonts,amssymb}
\usepackage{bm}
\usepackage{graphicx}
\usepackage{tabularx}
\usepackage{enumerate}
\usepackage{multirow}
\usepackage{booktabs}
\usepackage[font=footnotesize,labelfont=bf]{caption}
\usepackage{subfig}
\usepackage[svgnames,table]{xcolor}
\usepackage{latexsym}
\usepackage{amsmath}
\usepackage{booktabs}
\usepackage{multirow}
\usepackage{setspace}
\usepackage[toc,page]{appendix}
\usepackage{mathtools}
\usepackage{cancel}
\usepackage{relsize}
\usepackage{empheq}
\usepackage{mathrsfs}
\usepackage{breqn}
\usepackage{arcs}
\usepackage{amssymb}
\usepackage{wasysym}
\usepackage{pifont}

\begin{document}

\newcolumntype{L}[1]{>{\raggedright\let\newline\\\arraybackslash\hspace{0pt}}m{#1}}
\newcolumntype{C}[1]{>{\centering\let\newline\\\arraybackslash\hspace{0pt}}m{#1}}
\newcolumntype{R}[1]{>{\raggedleft\let\newline\\\arraybackslash\hspace{0pt}}m{#1}}

\def\ds{\displaystyle}

\newcommand{\beq}{\begin{equation}}
\newcommand{\eeq}{\end{equation}}
\newcommand{\lb}{\label}
\newcommand{\beqar}{\begin{eqnarray}}
\newcommand{\eeqar}{\end{eqnarray}}
\newcommand{\barr}{\begin{array}}
\newcommand{\earr}{\end{array}}
\newcommand{\jump}{\parallel}

\def\c{{\circ}}

\newcommand{\Ehat}{\hat{E}}
\newcommand{\That}{\hat{\bf T}}
\newcommand{\Ahat}{\hat{A}}
\newcommand{\chat}{\hat{c}}
\newcommand{\shat}{\hat{s}}
\newcommand{\khat}{\hat{k}}
\newcommand{\muhat}{\hat{\mu}}
\newcommand{\mc}{M^{\scriptscriptstyle C}}
\newcommand{\mei}{M^{\scriptscriptstyle M,EI}}
\newcommand{\mec}{M^{\scriptscriptstyle M,EC}}
\newcommand{\hbeta}{{\hat{\beta}}}
\newcommand{\rec}[2]{\left( #1 #2 \ds{\frac{1}{#1}}\right)}
\newcommand{\rep}[2]{\left( {#1}^2 #2 \ds{\frac{1}{{#1}^2}}\right)}
\newcommand{\derp}[2]{\ds{\frac {\partial #1}{\partial #2}}}
\newcommand{\derpn}[3]{\ds{\frac {\partial^{#3}#1}{\partial #2^{#3}}}}
\newcommand{\dert}[2]{\ds{\frac {d #1}{d #2}}}
\newcommand{\dertn}[3]{\ds{\frac {d^{#3} #1}{d #2^{#3}}}}

\def\bob{{\, \underline{\overline{\otimes}} \,}}
\def\ob{{\, \underline{\otimes} \,}}
\def\scalp{\mbox{\boldmath$\, \cdot \, $}}
\def\gdp{\makebox{\raisebox{-.215ex}{$\Box$}\hspace{-.778em}$\times$}}
\def\daa{\makebox{\raisebox{-.050ex}{$-$}\hspace{-.550em}$: ~$}}
\def\mK{\mbox{${\mathcal{K}}$}}
\def\cK{\mbox{${\mathbb {K}}$}}

\DeclarePairedDelimiter{\abso}{\lvert}{\rvert}
\DeclarePairedDelimiter{\norma}{\lVert}{\rVert}

\def\Xint#1{\mathchoice
   {\XXint\displaystyle\textstyle{#1}}%
   {\XXint\textstyle\scriptstyle{#1}}%
   {\XXint\scriptstyle\scriptscriptstyle{#1}}%
   {\XXint\scriptscriptstyle\scriptscriptstyle{#1}}%
   \!\int}
\def\XXint#1#2#3{{\setbox0=\hbox{$#1{#2#3}{\int}$}
     \vcenter{\hbox{$#2#3$}}\kern-.5\wd0}}
\def\ddashint{\Xint=}
\def\fpint{\Xint=}
\def\dashint{\Xint-}
\def\cpvint{\Xint-}
\def\intl{\int\limits}
\def\cpvintl{\cpvint\limits}
\def\fpintl{\fpint\limits}
\def\ointl{\oint\limits}
\def\bA{{\bf A}}
\def\ba{{\bf a}}
\def\bB{{\bf B}}
\def\bb{{\bf b}}
\def\bc{{\bf c}}
\def\bC{{\bf C}}
\def\bD{{\bf D}}
\def\bE{{\bf E}}
\def\be{{\bf e}}
\def\bbf{{\bf f}}
\def\bF{{\bf F}}
\def\bG{{\bf G}}
\def\bg{{\bf g}}
\def\bi{{\bf i}}
\def\bH{{\bf H}}
\def\bK{{\bf K}}
\def\bL{{\bf L}}
\def\bM{{\bf M}}
\def\bN{{\bf N}}
\def\bn{{\bf n}}
\def\b0{{\bf 0}}
\def\bo{{\bf o}}
\def\bX{{\bf X}}
\def\bx{{\bf x}}
\def\bP{{\bf P}}
\def\bp{{\bf p}}
\def\bQ{{\bf Q}}
\def\bq{{\bf q}}
\def\bR{{\bf R}}
\def\bS{{\bf S}}
\def\bs{{\bf s}}
\def\bT{{\bf T}}
\def\bt{{\bf t}}
\def\bU{{\bf U}}
\def\bu{{\bf u}}
\def\bv{{\bf v}}
\def\bw{{\bf w}}
\def\bW{{\bf W}}
\def\by{{\bf y}}
\def\bz{{\bf z}}
\def\T{{\bf T}}
\def\Te{\textrm{T}}
\def\Id{{\bf I}}
\def\bxi{\mbox{\boldmath${\xi}$}}
\def\balpha{\mbox{\boldmath${\alpha}$}}
\def\bbeta{\mbox{\boldmath${\beta}$}}
\def\bepsilon{\mbox{\boldmath${\epsilon}$}}
\def\bvarepsilon{\mbox{\boldmath${\varepsilon}$}}
\def\bomega{\mbox{\boldmath${\omega}$}}
\def\bphi{\mbox{\boldmath${\phi}$}}
\def\bsigma{\mbox{\boldmath${\sigma}$}}
\def\bfeta{\mbox{\boldmath${\eta}$}}
\def\bDelta{\mbox{\boldmath${\Delta}$}}
\def\btau{\mbox{\boldmath $\tau$}}
\def\tr{{\rm tr}}
\def\dev{{\rm dev}}
\def\div{{\rm div}}
\def\Div{{\rm Div}}
\def\Grad{{\rm Grad}}
\def\grad{{\rm grad}}
\def\Lin{{\rm Lin}}
\def\Sym{{\rm Sym}}
\def\Skw{{\rm Skew}}
\def\abs{{\rm abs}}
\def\Re{{\rm Re}}
\def\Im{{\rm Im}}
\def\capB{\mbox{\boldmath${\mathsf B}$}}
\def\capC{\mbox{\boldmath${\mathsf C}$}}
\def\capD{\mbox{\boldmath${\mathsf D}$}}
\def\capE{\mbox{\boldmath${\mathsf E}$}}
\def\capG{\mbox{\boldmath${\mathsf G}$}}
\def\tcapG{\tilde{\capG}}
\def\capH{\mbox{\boldmath${\mathsf H}$}}
\def\capK{\mbox{\boldmath${\mathsf K}$}}
\def\capL{\mbox{\boldmath${\mathsf L}$}}
\def\capM{\mbox{\boldmath${\mathsf M}$}}
\def\capR{\mbox{\boldmath${\mathsf R}$}}
\def\capW{\mbox{\boldmath${\mathsf W}$}}

\def\i{\mbox{${\mathrm i}$}}
\def\mC{\mbox{\boldmath${\mathcal C}$}}
\def\mB{\mbox{${\mathcal B}$}}
\def\mE{\mbox{${\mathcal{E}}$}}
\def\mL{\mbox{${\mathcal{L}}$}}
\def\mK{\mbox{${\mathcal{K}}$}}
\def\mV{\mbox{${\mathcal{V}}$}}
\def\C{\mbox{\boldmath${\mathcal C}$}}
\def\E{\mbox{\boldmath${\mathcal E}$}}

\def\ACME{{ Arch. Comput. Meth. Engng.\ }}
\def\ARMA{{ Arch. Rat. Mech. Analysis\ }}
\def\AMR{{ Appl. Mech. Rev.\ }}
\def\ASCEEM{{ ASCE J. Eng. Mech.\ }}
\def\acta{{ Acta Mater. \ }}
\def\CMAME {{ Comput. Meth. Appl. Mech. Engrg.\ }}
\def\CRAS{{ C. R. Acad. Sci., Paris\ }}
\def\EFM{{ Eng. Fract. Mech.\ }}
\def\EJMA{{ Eur.~J.~Mechanics-A/Solids\ }}
\def\IJES{{ Int. J. Eng. Sci.\ }}
\def\IJF{{ Int. J. Fracture\ }}
\def\IJMS{{ Int. J. Mech. Sci.\ }}
\def\IJNAMG{{ Int. J. Numer. Anal. Meth. Geomech.\ }}
\def\IJP{{ Int. J. Plasticity\ }}
\def\IJSS{{ Int. J. Solids Structures\ }}
\def\IngA{{ Ing. Archiv\ }}
\def\JAM{{ J. Appl. Mech.\ }}
\def\JAP{{ J. Appl. Phys.\ }}
\def\JE{{ J. Elasticity\ }}
\def\JM{{ J. de M\'ecanique\ }}
\def\JMPS{{ J. Mech. Phys. Solids\ }}
\def\JoMMS{{ J. Mech. Materials Structures\ }}
\def\Macro{{ Macromolecules\ }}
\def\MOM{{ Mech. Materials\ }}
\def\MMS{{ Math. Mech. Solids\ }}
\def\MMT{{ Metall. Mater. Trans. A}}
\def\MPCPS{{ Math. Proc. Camb. Phil. Soc.\ }}
\def\MRC{{ Mech. Res. Comm.}}
\def\MSE{{ Mater. Sci. Eng.}}
\def\PMPS{{ Proc. Math. Phys. Soc.\ }}
\def\PRE{{ Phys. Rev. E\ }}
\def\PRL{{ Phys. Rev. Letters\ }}
\def\PRSL{{ Proc. R. Soc.\ }}
\def\rock{{ Rock Mech. and Rock Eng.\ }}
\def\QAM{{ Quart. Appl. Math.\ }}
\def\QJMAM{{ Quart. J. Mech. Appl. Math.\ }}
\def\SCRMAT{{ Scripta Mater.\ }}
\def\SM{{\it Scripta Metall. }}

\def\salto#1#2{
[\mbox{\hspace{-#1em}}[#2]\mbox{\hspace{-#1em}}]}

\title{Hypocycloidal inclusions in nonuniform out-of-plane elasticity: stress singularity vs stress reduction}\date{}

\author{S. Shahzad, F. Dal Corso and D. Bigoni \\
DICAM, University of Trento, via Mesiano 77, I-38123 Trento, Italy }

\maketitle

\begin{abstract}
\noindent
Stress field solutions and Stress Intensity Factors (SIFs) are found for $n$-cusped hypocycloidal shaped voids and rigid inclusions
in an infinite linear elastic plane subject to nonuniform remote antiplane loading, using complex
potential and conformal mapping.
It is shown that a void with
hypocycloidal shape can lead
to a higher SIF than that induced by a corresponding star-shaped crack; this is counter intuitive as the latter usually produces a more severe stress field in the material.
Moreover, it
is observed that when the order $m$ of the polynomial governing the
remote loading grows, the stress fields generated by the
hypocycloidal-shaped void and the star-shaped crack tend to
coincide, so that they become equivalent from the point of view of a
failure analysis. Finally, special geometries and loading conditions
are discovered for which there is no stress singularity at the
inclusion cusps and where the stress is even reduced with respect to
the case of the absence of the inclusion. The concept of Stress
Reduction Factor (SRF) in the presence of a sharp wedge is therefore
introduced, contrasting with the well-known definition of Stress
Concentration Factor (SCF) in the presence of inclusions with smooth
boundary.
The results presented in this paper provide criteria that will help in the design of ultra strong composite materials, where stress singularities always promote failure.
Furthermore, they will
facilitate finding the special conditions where resistance can be optimized in the presence of inclusions with non-smooth boundary.

\end{abstract}


\section{Introduction}

The determination of the stress field near a crack, a stiffener, an
inclusion, or a defect in an elastic matrix material is a key
problem in the design of composites \cite{ammari, ammari2, barbieri,
craciun, dascalu, panos, roaz, radi}, as such stress fields exhibit
strong stress concentrations that can impose severe limitations on
the strength of composites.

In the present article a rigid inclusion or a void is analyzed with
a \emph{hypocycloidal} shape of order $n$ embedded in an elastic-isotropic plane subject to a remote loading condition of \emph{nonuniform} antiplane shear represented as a polynomial of order $m$.
For uniform remote load ($m=0$), this problem has been thoroughly investigated,
for both the cases of rigid inclusions and voids, when plane \cite{chen, chen2,
 hasebe_multiple_crack, chu, gao, gdoutos, movmov, trush, pana2, pana3} or antiplane \cite{movchan, long, vigder1, vigder2}
 conditions prevail; a case of nonlinear elastic behaviour has also been recently considered \cite{wang}.
However, the disuniformity in the applied load ($m\neq 0$), analyzed here for the first time, yields unexpected and counter intuitive results, which are
important to understand the complexity arising form the highly-varying fields that can develop in composite materials deformed in extreme conditions.
In particular,
in the present work polynomial \emph{shear} loading at infinity is
prescribed (as in \cite{partI, partII, vasudevan}) to an elastic-isotropic plane, containing
a void or a rigid inclusion with an $n$-cusped hypocycloidal shape.
This problem is solved
in an analytical form both for the stress full-field and for the stress intensity factor, solutions which reveal several phenomena of
interplay between stress singularities and stress reduction, which remain undetected for uniform applied loads.
The most important of these phenomena are the following.

(i.) Under uniform stress conditions the stress intensity factor at a cusp of an hypocycloidal void is always smaller
than that at a tip of a star-shaped crack, so that a crack tip is more detrimental to strength than a cusp, the same is not true for certain orders $m$ of polynomial fields of remote loading. In these cases, a hypocycloidal void leads a material to failure more easily than a star-shaped crack.

(ii.) While for uniform loading a stress singularity is always present at a cusp, for certain orders $m$ of polynomial fields of remote loading, this
singularity can disappear, so that the stress can lie below the value corresponding to the
unperturbed field.
This effect has been quantified by introducing the notion of \lq Stress Reduction Factor' (SRF), which is shown to increase with the number $n$ of cusps
and to decrease with the order $m$ of the polynomial load. Note that the concept of SRF contrasts with the well-known definition
of Stress Concentration
Factor, introduced in the presence of inclusions with smooth boundaries and representing the increase of the stress state
at the inclusion boundary with respect to the case when the inclusion is absent.

(iii.) For orders of polynomial loading $m$ much greater than the number $n$ of cusps of the hypocycloidal inclusion,
the stress state generated in the matrix tends to coincide with that generated by an $n$-pointed star-shaped crack or rigid inclusion.

\vspace{3 mm}

\begin{figure}[!htb]
\begin{center}
\includegraphics[width=12 cm]{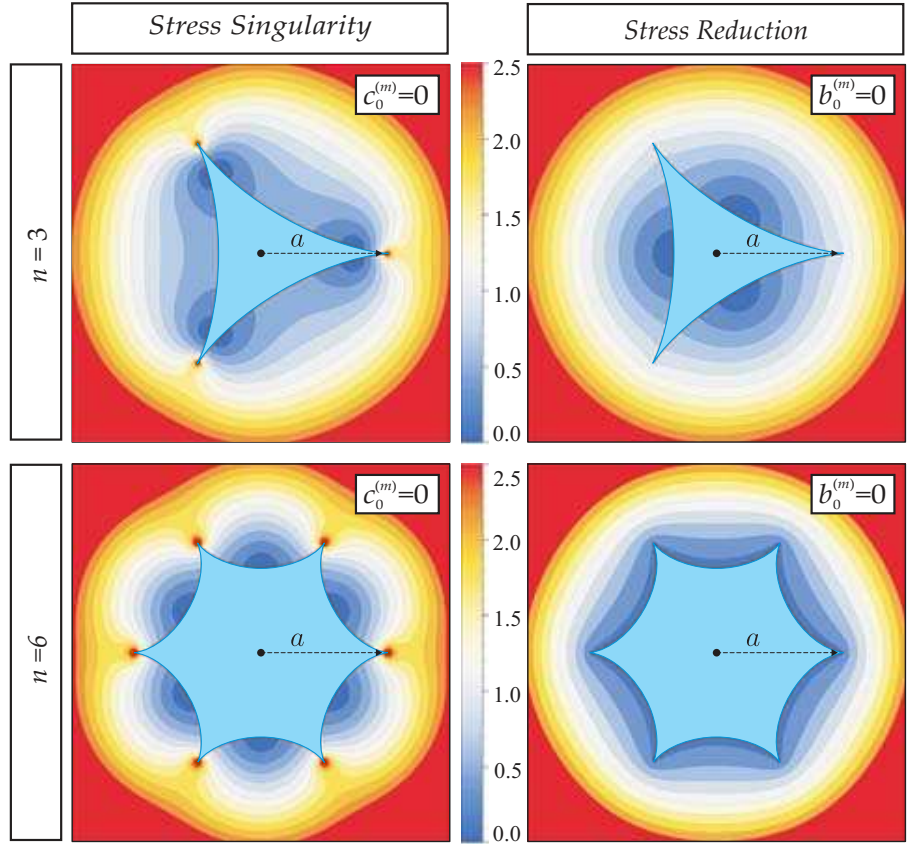}
\caption{
Level sets of dimensionless shear stress modulus $\tau^{(m)}(x_1, x_2)/\tau^{\infty(m)}(a,0)$
near $n$-cusped rigid inclusions ($n=3$ and $n=6$), subject to quadratic ($m=2$) remote out-of-plane shearing,
showing
stress singularity (left column, $c_{0}^{(m)}=0$) and stress reduction (right column, $b_{0}^{(m)}=0$).
In the figures a stress singularity corresponds to the appearance of
a \lq red spot' at the inclusion cusp.
In the cases showing stress reduction, the
stress singularity is absent and the stress at all the cusps is smaller than that which would be attained without the inclusion.
}
\label{fig_stress_reduction_fullfield_and_curve_1}
\end{center}
\end{figure}

Some of the above concepts are elucidated in Fig. \ref{fig_stress_reduction_fullfield_and_curve_1}, where the level sets of
 the modulus of the shear stress are plotted from the closed form solution (obtained in Section \ref{baaaazzzooo})
near three- and six- cusped rigid hypocycloidal inclusions,
for quadratic ($m=2$) loadings. The \lq red spots' visible near the inclusion cusps in Fig. \ref{fig_stress_reduction_fullfield_and_curve_1} (left column)
are the signature of stress singularity
generated when the loading is defined by $c_0^{(m)}=0$.
When the loading is defined by $b_0^{(m)}=0$ stress singularities disappear, so that stress reduction occurs at all the vertices of
the inclusion (Fig. \ref{fig_stress_reduction_fullfield_and_curve_1},
right column), where the stress falls below the value that would be present
if the inclusion was absent (see also Fig. \ref{fig_stress_reduction_fullfield_and_curve_2}).

Photoelastic investigations  \cite{inclusioni, noselli} confirm the
severe stress fields theoretically predicted for materials
containing stiff inclusions and show that these fields may yield
failure instead of reinforcement for composite materials
\cite{stiffener0,hasebe,misra,ozturk}. Therefore, the results given
in the present article show possibilities of greatly enhancing the
strength of composites through the careful design of the inclusion
shapes, thus opening the way to the realization of ultra-strong
materials.

\section{Out-of-plane elasticity and nonuniform remote conditions} \lb{gogo}

A linear elastic solid with shear modulus $\mu$ is considered subject to out-of-plane conditions, so that the
only non-vanishing displacement is $w(x_1, x_2)$ in the $x_3$ direction, orthogonal to the $x_1$--$x_2$ plane.
Constitutive equations lead to the following non-null shear stress $\tau_{\alpha 3}$ components ($\alpha=1,2$)
\beq\label{shear}
\tau_{13}=\mu \,w_{,1}, \qquad
\tau_{23}=\mu \,w_{,2},
\eeq
where a comma denotes differentiation with respect to the variables $x_1$ or $x_2$, so that, in the absence of body forces, equilibrium is expressed by
the Laplace equation for displacement field $w$
\beq\label{laplace}
\nabla^2 w=0.
\eeq

Following \cite{partI,partII,schiavone,vasudevan}, an infinite class of boundary conditions for out-of-plane problems is considered through
the following polynomial expression of $m$-th order ($m \in \mathbb{N}$) for the remote displacement applied to an infinite
elastic plane. Considering a polar coordinate system $(r, \theta)$ centered at the origin of the $x_1$--$x_2$ axes,
the polynomial displacement boundary conditions can be written as
    \beq
        \label{eq_general_b.c.polar2}
            w^{\infty (m)}(r, \theta)=\frac{r^{m+1}}{\mu (m+1)}
             \left[b^{(m)}_0 \cos\left((m+1)\theta\right)+c^{(m)}_0 \sin\left((m+1)\theta\right)\right],
    \eeq
where $b^{(m)}_{0}$ and $c^{(m)}_{0}$ are two constants defining the boundary condition for each order $m$ of the polynomial.
Polynomial displacement boundary conditions is a key tool in problems of homogenization for higher-order continua,
as a representation of the expansion of the displacement field near a point \cite{bacca}.

The remote
displacement condition (\ref{eq_general_b.c.polar2}), through the linear constitutive relationship,
corresponds to the following stress components at infinity
    \beq
        \begin{cases}
        \barr{cll}
        \label{eq_general_b.c.polar}
            \tau_{r 3}^{\infty (m)}(r, \theta)&=&
            r^{m} \left[b^{(m)}_0 \cos\left((m+1)\theta\right)+c^{(m)}_0 \sin\left((m+1)\theta\right)\right],\\[6mm]
            \tau_{\theta 3 }^{\infty (m)}(r, \theta)&=&
            r^{m} \left[c^{(m)}_0 \cos\left((m+1)\theta\right)-b^{(m)}_0 \sin\left((m+1)\theta\right)\right].
                \earr
        \end{cases}
    \eeq
Considering the relationship between the Cartesian and the polar systems, the shear stress components (\ref{eq_general_b.c.polar}) can
be expressed as
    \beq
        \label{eq_general_b.c.}
            \tau_{13}^{\infty (m)}(x_{1},x_{2})=
            \mathlarger{\sum_{j=0}^{m}} b_j^{(m)} x_1^{m-j}x_2^{j},\qquad
                        \tau_{23}^{\infty (m)}(x_{1},x_{2})=
            \mathlarger{\sum_{j=0}^{m}} c_j^{(m)} x_1^{m-j}x_2^{j},
                \eeq
where the constants $b^{(m)}_{j}$ and $c^{(m)}_{j}$ ($j=1,..., m$) are related to the two coefficients
$b^{(m)}_{0}$ and $c^{(m)}_{0}$  through
\beq
\begin{array}{ccc}
b_j^{(m)}=\ds (-1)^{\lfloor j/2\rfloor}\binom{m}{j} \frac{b_0^{(m)}(1+(-1)^j)+c_0^{(m)}(1-(-1)^j)}{2},\\[4 mm]
c_j^{(m)}=\ds (-1)^{\lceil j/2\rceil}\binom{m}{j}  \frac{b_0^{(m)}(1-(-1)^j)+c_0^{(m)}(1+(-1)^j)}{2},
\end{array}
\eeq
where $\binom{\cdot}{\cdot\cdot}$ denotes the binomial coefficient, while the symbols $\lfloor \cdot \rfloor$ and $\lceil \cdot \rceil$
stand repectively for  the
largest integer less than or equal to and for the smallest integer greater than or equal to the relevant argument.

\section{Closed-form solution for an infinite elastic plane containing a void or rigid inclusion with
a hypocycloidal shape}\lb{baaaazzzooo}

In this Section, the closed-form solution is obtained for the out-of-plane problem of a $n$-cusped hypocycloidal void or rigid inclusion embedded in an infinite elastic material with remote
boundary conditions expressed by the polynomial (\ref{eq_general_b.c.polar2}).
The two-dimensional Laplace equation (\ref{laplace}) for the displacement field $w$ can be solved through the complex potential technique
where the complex variable $z=x_1+\i x_2$ is introduced.
The boundary of a generic inclusion in an infinite plane can be mapped into a circle of unit radius in the
conformal plane (where the position is given by the variable $\zeta$) by means of a conformal mapping $z=\omega(\zeta)$.
The out-of-plane elastic problem can therefore be solved through a complex potential representation $g(\zeta)$, related to the displacement $w$,
stresses $\tau_{13}$, $\tau_{23}$,
and  shear force resultant $F_{\stackrel\frown {B C}}$ (along the arc $\stackrel\frown {B C}$) as
\beq
\label{eq_transformed_fullfield}
w=\frac{1}{\mu}\Re[g(\zeta)], \qquad \tau_{13}-\i \tau_{23}=\frac{g'(\zeta)}{\omega'(\zeta)},
\qquad
F_{\stackrel\frown {BC}}=\Im \left[g(\zeta_{B})-g(\zeta_{C})\right].
\eeq

In the case of hypocycloidal inclusions, with a number $n$ of cusps ($n \in \mathbb{Z} $, $n \geq 2$),
the function $\omega(\zeta)$  mapping
 the exterior region of the inclusion (within the physical $z-plane$) onto the exterior region of the unit circle
(within the conformal $\zeta-plane$, see the inset in Fig. \ref{fig_sif_hypo}) is given by \cite{ivanov}
 \beq
 \label{map_cusp}
 \barr{cll}
 \omega(\zeta)&=&\ds a \Omega \left(\zeta+\frac{1}{n-1}\zeta^{1-n}\right),
 \earr
 \eeq
where $a$ is the radius of the circle inscribing the inclusion and $\Omega$ is the scaling factor of the inclusion,
 function of the cusp number $n$ as
\beq
\label{omega_cusp}
\Omega(n)= \ds \frac{n-1}{n} \quad \in\,\, \left[\frac{1}{2}\,,1\right).
\eeq
Note that the conformal mapping for the hypocycloidal inclusion (\ref{map_cusp}) provides the well-known conformal mappings for
line inclusion (crack or stiffener) of length $2a$, in the case $n=2$, and for a circle of radius $a$, in the limit  $n\rightarrow \infty$,  which are respectively given by
 \beq
  \omega(\zeta)=\ds \frac{a}{2} \left(\zeta+\frac{1}{\zeta}\right),
 \qquad
\omega(\zeta)=\ds a \zeta.
 \eeq

Applying the superposition principle, the complex potential $g(z)$ is the sum of the unperturbed $g^{\infty}(z)$ and
perturbed $g^{p}(z)$ potentials, the former describing the solution in the case that the inclusion is absent
while the latter defining the perturbation introduced by the presence of the inclusion
\beq\label{somma_g}
g(\zeta)=g^{\infty}(\zeta)+g^{p}(\zeta).
\eeq
With reference to the polynomial expression (\ref{eq_general_b.c.})
for the remote displacement boundary condition, the unperturbed potential can be rewritten as
\beq
\label{eq_unperturbed_potential}
g^{\infty}(\zeta)= T^{(m)} \left[\omega(\zeta)\right]^{m+1} \, ,
\eeq
where $T^{(m)}$ is the following function of the loading parameters $b_0^{(m)}$ and $c_0^{(m)}$
\beq
\label{eq_unperturbed_potential_1}
T^{(m)}= \frac{b^{(m)}_0-\i  \, c^{(m)}_0}{m+1}.
\eeq

By means of the  binomial theorem, the unperturbed potential (\ref{eq_unperturbed_potential}) can be expressed  as
\beq
\label{eq_unperturbed_t_reali_asterisk}
g^{\infty} (\zeta)= \left(a \Omega(n) \right)^{m+1} \,\,T^{(m)}\sum_{j=0}^{m+1} \binom{m+1}{j} \, \frac{\zeta^{m+1-j n}}{(n-1)^j}.
\eeq

Considering that the null traction resultant condition $F_{\stackrel{\frown}{BC}}=0$ holds
for the  hypocycloidal void ($\chi=1$), while the rigid-body displacement condition
$w_B=w_C$ holds for (every pair of points $B$ and $C$ along the boundary of) the hypocycloidal rigid inclusion ($\chi=-1$),
the perturbed complex potential $g^{p}$ is obtained in the form
\begin{dmath}
\label{gp_cusp}
g^{p} (\zeta)=\left(a \Omega(n) \right)^{m+1} \left\{\chi\overline{T^{(m)}} \left[-\frac{(q n)!}{q![q (n-1)]!(n-1)^q}\delta_{m+1,qn}+
\sum_{j=0}^{q} \binom{m+1}{j} \, \frac{1}{(n-1)^j \,\zeta^{m+1-j n}}
\right]-
T^{(m)}\sum_{j=q+1}^{m+1} \binom{m+1}{j} \, \frac{1}{(n-1)^j \,\zeta^{j n-m-1}}\right\},
\end{dmath}
where the integer parameter $q=\left\lfloor (m+1)/n\right\rfloor$ is introduced,
so that the complex potential $g(\zeta)$ follows from eqn (\ref{somma_g}) as
\begin{dmath}
\label{gtot_cusp}
g(\zeta)=\left(a \Omega(n) \right)^{m+1} \left[-\frac{(q n)!}{q![q (n-1)]!(n-1)^q}\chi\overline{T^{(m)}}\delta_{m+1,qn}+
 \sum_{j=0}^{q} \binom{m+1}{j} \frac{1}{(n-1)^j}\left(T^{(m)}\zeta^{m+1-j n}
+\frac{\chi \overline{T^{(m)}}}{\zeta^{m+1-j n}}\right) \right].
\end{dmath}

It is worth noting that solution (\ref{gtot_cusp})  simplifies in some special cases, as
\begin{itemize}
\item
$n>m+1$ (or, equivalently, $q=0$)
\beq
\label{eq_total_potential_poly_star_crack_case3_tot}
g(\zeta)=\left(a \Omega(n) \right)^{m+1} \left[T^{(m)} \zeta^{m+1} + \frac{\chi \overline{T^{(m)}}}{\zeta^{m+1}}\right],
\eeq
which is similar to the solution in the conformal plane for the circular and polygonal inclusions;
\item
$m=0$ (corresponding to the case of uniform antiplane shear \cite{hasebe_multiple_crack})
\beq
g(\zeta)= a \Omega(n) \left[T^{(0)} \zeta + \frac{\chi \overline{T^{(0)}}}{\zeta}\right].
\eeq
\end{itemize}

\section{Stress singularities and stress reduction} \lb{SIIIFFFF}

As in the cases of cracks and rigid line inclusions \cite{partI, partII, japan, sih}, a square root stress singularity at the cusps of an hypocycloidal void or rigid inclusion is predicted in the theory of elasticity \cite{hasebe_multiple_crack, long, gao, trush}.
In this section the Stress Intensity Factors are derived as functions of the cusp number $n$ and of the polynomial order $m$ of the load. Moreover,
conditions for which stress singularities disappear are defined and a special feature is observed, namely, the stress reduction, which corresponds to the fact that the stress measured at the cusp is smaller than that present at the same point in the absence of the inclusion.

From the solution (\ref{gtot_cusp}) obtained in the previous Section, the stress field near a $n$-cusped hypocycloidal void and rigid inclusion
is given by
\beq
\label{eq_tauconf_star_crack}
\tau_{13}^{(m)}-\i \tau_{23}^{(m)}= \frac{(a \Omega(n))^m}{\left(1-\zeta^{-n}\right)}
\ds \sum_{j=0}^{q}  \binom{m+1}{j} \frac{m+1-nj}{(n-1)^j}  \left(T^{(m)}\zeta^{m-jn} -\frac{\chi \overline{T^{(m)}}}{\zeta^{m+2-j n}}\right).
\eeq
Focusing the attention on the inclusion cusp  located at the point ($x_1 = a$, $x_2=0$) and introducing  a local reference system centered in the mapped cusp ($\zeta^*=\zeta-1$),
the stress fields   (\ref{eq_tauconf_star_crack}) can be expanded about the cusp in the limit of $|\zeta^*|\rightarrow 0$
(corresponding to the limit $z\rightarrow a$) as
\beq
 \tau_{13}^{(m)}-\i \tau_{23}^{(m)}  \approx   \frac{(a \Omega(n))^m}{n(m+1)\zeta^*}\ds \left[b^{(m)}_0(1-\chi)
-\i c^{(m)}_0(1+\chi)\right] \sum_{j=0}^{q} \binom{m+1}{j} \frac{(m+1-j n)}{(n-1)^j}.
\eeq

Expansion of the conformal mapping (\ref{map_cusp}) leads to the asymptotics between the
relative coordinate $z^*=\rho e^{\i\vartheta}$ and the conformal relative coordinate $\zeta^*$ in the form
\beq
\label{eq_inversa}
\zeta^* \simeq \ds \sqrt{\frac{2\rho}{a (n-1)}  }\,\, e^{\frac{\i\vartheta}{2}},
\eeq
so that the  shear stress components can be approximated around the cusp as
\begin{dmath}
\label{asym_cusp}
 \tau_{13}^{(m)}-\i \tau_{23}^{(m)} \simeq   \frac{1}{\sqrt{2 \rho}}\frac{[a(n-1)]^{m+\frac{1}{2}}}{ (m+1)n^{m+1}}
\left[b^{(m)}_0(1-\chi)
-\i c^{(m)}_0(1+\chi)\right] \left(\cos\frac{\vartheta}{2}-\i\, \sin\frac{\vartheta}{2}\right)\sum_{j=0}^{q} \binom{m+1}{j} \frac{(m+1-j n)}{(n-1)^j},
\end{dmath}
highlighting the square root singularity in the stress field at the cusp, as predicted by the asymptotics around sharp notch \cite{movmov}.

\subsection{Stress Intensity Factors}

Stress Intensity Factors (SIFs) for the symmetric and the
anti-symmetric out-of-plane problem are defined, respectively, as
follows \cite{radaj} \beq\label{sif} K_{\textup{III}}^{\textup{S}}=
\lim_{\rho \rightarrow 0} \sqrt{2\pi \rho} \,\,
\tau_{13}(\rho,0),\qquad K_{\textup{III}}^{\textup{A}}=\lim_{\rho
\rightarrow 0} \sqrt{2\pi \rho} \,\, \tau_{23}(\rho,0), \eeq so
that, considering the asymptotic stress field (\ref{asym_cusp}), the
expression of the SIFs for a $n$-cusped hypocycloidal inclusion can
be analytically obtained in the following closed-form expression
\beq \label{SIFSIF0}
\begin{bmatrix}
K_{\textup{III}}^{\textup{S}}(n,m)\\[3mm]
K_{\textup{III}}^{\textup{A}}(n,m)
\end{bmatrix}
=   \frac{\sqrt{\pi a} (n-1)^{m+\frac{1}{2}} a^m}{n^{m+1}(m+1)}
\left[\sum_{j=0}^{q} \binom{m+1}{j} \frac{(m+1-j n)}{(n-1)^j}\right]
\begin{bmatrix}
(1-\chi)b_0^{(m)}\\[3mm]
(1+\chi)c_0^{(m)}
\end{bmatrix}.
\eeq

Recalling that the definition (\ref{eq_general_b.c.}) for the remote applied shear stress implies
\beq\label{guardaunpo}
\tau_{13}^{\infty (m)}(x_{1}=a,0)= b_0^{(m)} a^{m},\qquad
\tau_{23}^{\infty (m)}(x_{1}=a,0)= c_0^{(m)} a^{m},
\eeq
the particularization of eq (\ref{SIFSIF0}) to the case of a void or a rigid n-cusped hypocycloidal inclusion,
leads to
\beq
\label{eq_sif_speciali_hypo}
\begin{bmatrix}
K_{\textup{III}}^{\text{{\tiny\ding{70}}}\textup{S}}(n,m)\\[3mm]
K_{\textup{III}}^{\text{{\tiny\ding{71}}}\textup{A}}(n,m)
\end{bmatrix}
=\frac{2\sqrt{\pi a} (n-1)^{m+\frac{1}{2}}}{n^{m+1}(m+1)}
\left[\sum_{j=0}^{q} \binom{m+1}{j} \frac{(m+1-j n)}{(n-1)^j}\right]
\left[
\begin{array}{ll}
\tau_{13}^{\infty(m)}(a,0)\\
\tau_{23}^{\infty(m)}(a,0)
\end{array}
\right],
\eeq
and
\beq
K_{\textup{III}}^{\text{{\tiny\ding{71}}}\textup{S}}(n,m)=K_{\textup{III}}^{\text{{\tiny\ding{70}}}\textup{A}}(n,m)=0,
\eeq
where the apexes $^{\text{\tiny\ding{71}}}$ and $^{\text{\tiny\ding{70}}}$ have been introduced to distinguish between the cases of voids and rigid inclusions.

\begin{figure}[!h]
  \begin{center}
\includegraphics[width=12 cm]{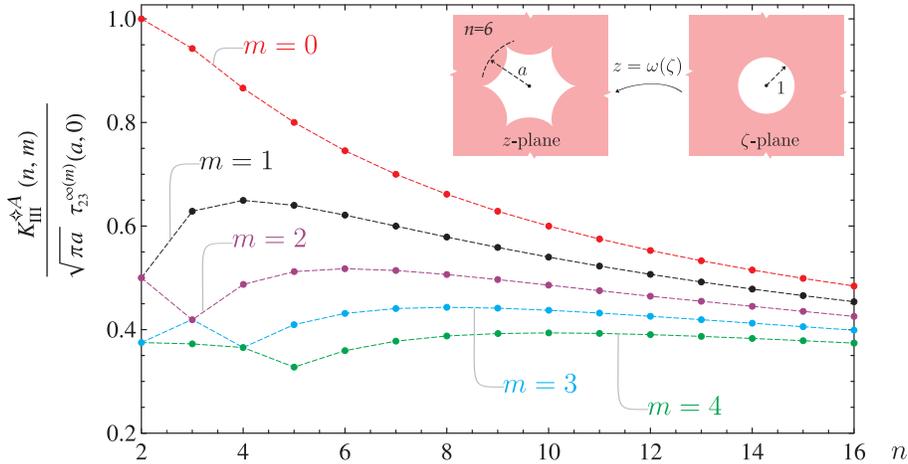}
\caption{ Stress Intensity Factors for $n$-cusped hypocycloidal voids
${K}_{\textup{III}}^{\text{{\tiny\ding{71}}}A}$, eq (\ref{eq_sif_speciali_hypo}), normalized through division
by the unperturbed stress evaluated at the inclusion cusp $\tau_{23}^\infty$($a,0$), as a
 function of the order $m$ of the remote loading. Note that the curves also represent
the Stress Intensity Factors for $n$-cusped hypocycloidal rigid inclusions ${K}_{\textup{III}}^{\text{{\tiny\ding{70}}}S}$,
 when normalized through division by the unperturbed stress evaluated at the inclusion cusp $\tau_{13}^\infty$($a,0$).
The inset shows the conformal mapping transforming the exterior region of a hypocycloidal inclusion (inscribed in a circle of radius $a$)
into the exterior region of a circular inclusion of unit radius in the $\zeta$-plane.
}
\label{fig_sif_hypo}
 \end{center}
\end{figure}

Expression (\ref{eq_sif_speciali_hypo}) for the SIFs simplifies in the special case $n>m+1$ (and therefore $q=0$)
\beq
\label{eq_sif_speciali2}
\begin{bmatrix}
K_{\textup{III}}^{\text{{\tiny\ding{70}}}\textup{S}}(n,m)\\[3mm]
K_{\textup{III}}^{\text{{\tiny\ding{71}}}\textup{A}}(n,m)
\end{bmatrix}
=\frac{2\sqrt{\pi a} (n-1)^{m+\frac{1}{2}}}{n^{m+1}}
\left[
\begin{array}{ll}
\tau_{13}^{\infty(m)}(a,0)\\[3mm]
\tau_{23}^{\infty(m)}(a,0)
\end{array}
\right] .
\eeq

The above case (\ref{eq_sif_speciali2}) embraces an infinite set of solutions, one of such solutions is that for an astroid ($n=4$) subject to uniform, linear and quadratic remote out-of-plane shear load ($m=0,1,2$).
Moreover, expression (\ref{eq_sif_speciali_hypo}) reduces to the values obtained for SIFs in the case of cracks or stiffeners ($n=2$)
\beq
\label{eq_sif_speciali_HYPO}
\begin{bmatrix}
K_{\textup{III}}^{\text{{\tiny\ding{70}}}\textup{S}}(n=2,m)\\[3mm]
K_{\textup{III}}^{\text{{\tiny\ding{71}}}\textup{A}}(n=2,m)
\end{bmatrix}
=\frac{\sqrt{\pi a}}{2^{m}(m+1)}
\left[\sum_{j=0}^{q} \binom{m+1}{j}  (m+1-2 j) \right]
\left[
\begin{array}{ll}
\tau_{13}^{\infty(m)}(a,0)\\[3mm]
\tau_{23}^{\infty(m)}(a,0)
\end{array}
\right]
.
\eeq

Note that, in the particular case of uniform antiplane shear ($m=0$), equation (\ref{eq_sif_speciali_hypo})
provides the same result as equation (4.142) in \cite{hasebe_multiple_crack}.

\begin{figure}[!h]
  \begin{center}
\includegraphics[width=16 cm]{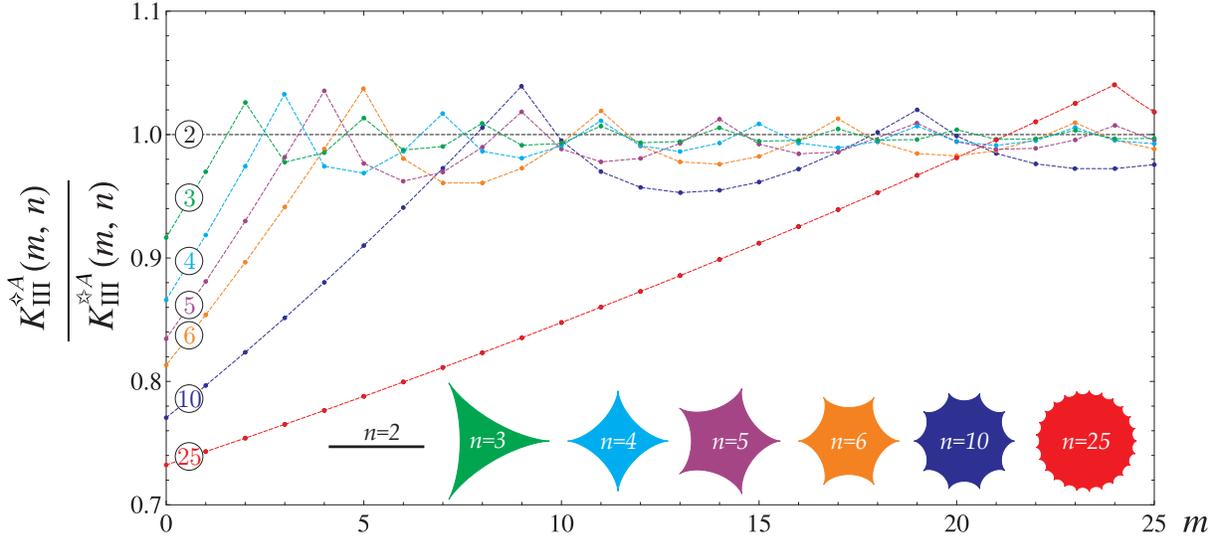}
\caption{ Ratio between the Stress Intensity Factors for
 an $n$-cusped hypocycloidal void
${K}_{\textup{III}}^{\text{{\tiny\ding{71}}}\textup{A}}$, eqn (\ref{eq_sif_speciali_HYPO}), and that for an $n$-pointed star-shaped crack
${K}_{\textup{III}}^{\text{{\tiny\ding{73}}}\textup{A}}$, eqn (\ref{eq_sif_speciali}), as
 a function of the order $m$ of the remote loading. A mechanical equivalence for the two geometries is observed
 when $m\gg n$.
  Note that
  the curves also represent
the Stress Intensity Factors ratio for an $n$-cusped hypocycloidal rigid inclusion ${K}_{\textup{III}}^{\text{{\tiny\ding{70}}}\textup{S}}$
and for an $n$-pointed star-shaped stiffener ${K}_{\textup{III}}^{\text{{\tiny\ding{72}}}\textup{S}}$.
}
\label{fig_sif_crack_hypo}
 \end{center}
\end{figure}

\begin{figure}[!htb]
  \begin{center}
\includegraphics[width=15 cm]{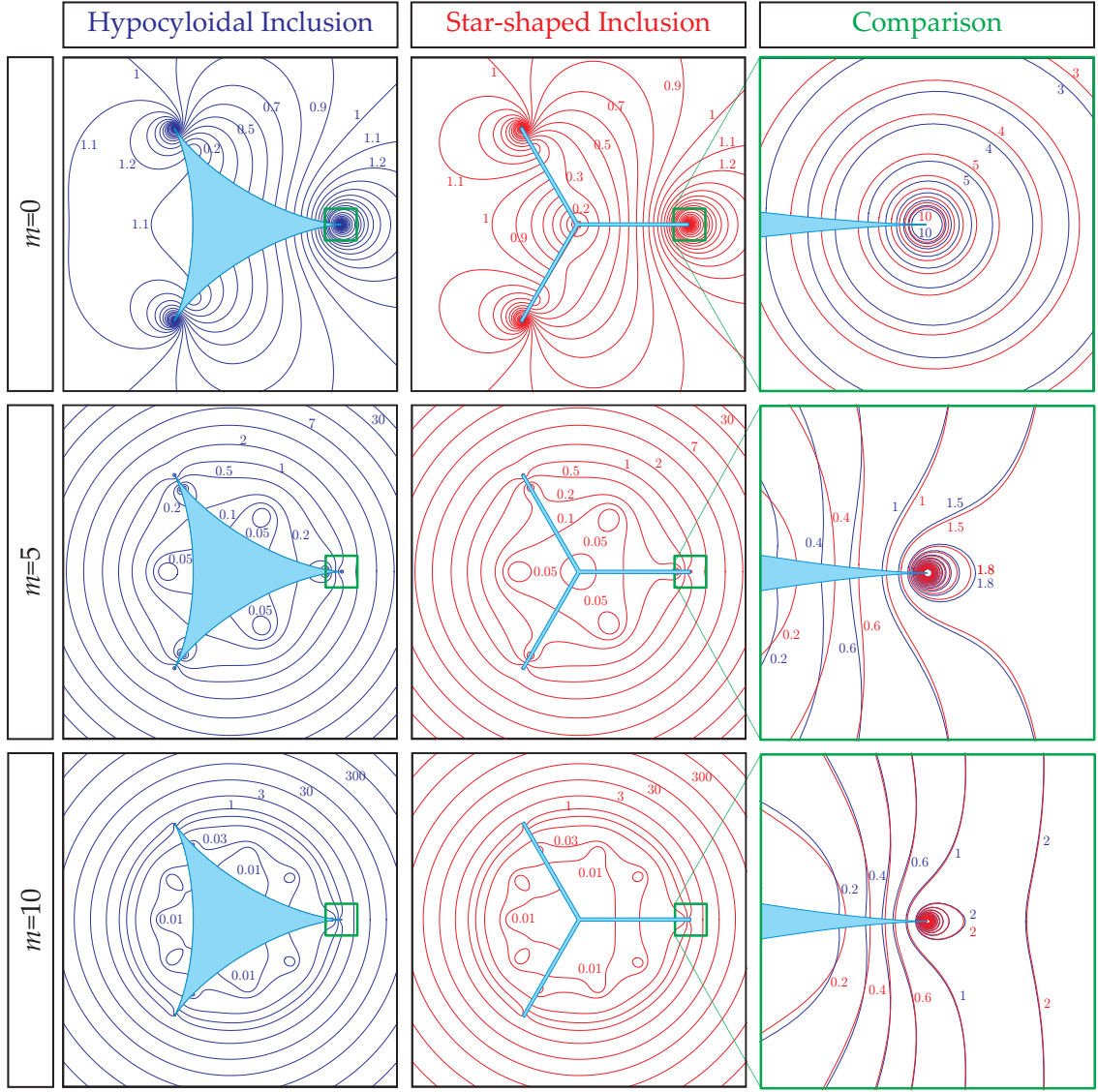}
\caption{Level sets of dimensionless shear stress modulus $\tau^{(m)}(x_1, x_2)/\tau^{\infty(m)}(a,0)$ highlighting
stress singularities near $n$-cusped rigid inclusions (left column) and $n$-pointed star-shaped stiffeners (central column)
at different orders $m$ of shear load ($c_{0}^{(m)}=0$). The right column shows comparisons between the two stress fields
reported in the left and central columns near the inclusion vertex having coordinates $x_1=a$ and $x_2=0$.
The mechanical equivalence can be noted between the two
inclusion geometries  for $m \gg n$. Note that these level sets also represent
the stress state near an $n$-cusped hypocycloidal void (left column)
and an $n$-pointed star-shaped crack (central column) in the case when the loading is provided by $b_{0}^{(m)}=0$.}
\label{fig_crack_hypo_fullfield_comparison_curve}
 \end{center}
\end{figure}

The SIFs for a $n$-cusped hypocycloidal inclusion (\ref{eq_sif_speciali_hypo}) satisfy the following properties
\beq
\frac{K_{\textup{III}}^{\text{{\tiny\ding{71}}}\textup{A}}(n,m)}{\tau^{\infty(m)}_{23}(a,0)}\geq
\frac{K_{\textup{III}}^{\text{{\tiny\ding{71}}}\textup{A}}(n,m+1)}{\tau^{\infty(m+1)}_{23}(a,0)}, ~~~
\frac{K_{\textup{III}}^{\text{{\tiny\ding{70}}}\textup{S}}(n,m)}{\tau^{\infty(m)}_{13}(a,0)}\geq
\frac{K_{\textup{III}}^{\text{{\tiny\ding{70}}}\textup{S}}(n,m+1)}{\tau^{\infty(m+1)}_{13}(a,0)},
\eeq
and are reported in Fig. \ref{fig_sif_hypo} for different values of $n$ and $m$. These values are also reported
in Fig. \ref{fig_sif_crack_hypo}
through normalization with the respective values for $n$-pointed star-shaped inclusions \cite{partII}
\beq
\label{eq_sif_speciali}
\begin{bmatrix}
K_{\textup{III}}^{\text{{\tiny\ding{72}}}\textup{S}}(n,m)\\[3mm]
K_{\textup{III}}^{\text{{\tiny\ding{73}}}\textup{A}}(n,m)
\end{bmatrix}
=\frac{2^{\frac{3n-4(m+1)}{2n}}\sqrt{\pi a}}{(m+1)\sqrt{n}}
\left[\sum_{j=0}^{q} \frac{(m+1-j n)}{j!} \prod_{l=0}^{j-1} \left(\frac{2(m+1)}{n}-l\right)\right]
\left[
\begin{array}{ll}
\tau_{13}^{\infty(m)}(a,0)\\
\tau_{23}^{\infty(m)}(a,0)
\end{array}
\right].
\eeq
It can be noted from Fig. \ref{fig_sif_crack_hypo} that:
\begin{itemize}
\item differently from the uniform loading case ($m=0$), the stress intensification around a cusp can be higher than that
occurring around a crack, so that a cusp can be more detrimental to failure than a crack for certain values of $m$. Moreover, the following relations have been  found numerically (an analytical proof looks awkward) to hold for every value used for $n$ and $m$
\beq
K^{\text{\tiny\ding{71}}\textup{A}}_{\textup{III}}(n,m=n-1)
\geq
K^{\text{\tiny\ding{73}}\textup{A}}_{\textup{III}}(n,m=n-1),
\qquad
K^{\text{\tiny\ding{70}}\textup{S}}_{\textup{III}}(n,m=n-1)
\geq
K^{\text{\tiny\ding{72}}\textup{S}}_{\textup{III}}(n,m=n-1).
\eeq
\item While the ratio between the SIFs for a hypocycloidal void and a star-shaped crack displays a monotonic increase for $m<n-1$,
an oscillatory behaviour around 1 is observed for $m>n-1$.  Such an  oscillation in the SIFs' ratio evidences a
decreasing amplitude, with peaks corresponding to the values $m=n j -1$ ($j \in \mathbb{N}_1$).
Therefore, when $m\gg n$, the
SIF for the hypocycloidal inclusion approaches that for star-shaped inclusion, namely, the following relations have been numerically found to hold
\beq
K^{\text{\tiny\ding{73}}\textup{A}}_{\textup{III}}(n,m)\simeq K^{\text{\tiny\ding{71}}\textup{A}}_{\textup{III}}(n,m),
\qquad
K^{\text{\tiny\ding{72}}\textup{S}}_{\textup{III}}(n,m)\simeq K^{\text{\tiny\ding{70}}\textup{S}}_{\textup{III}}(n,m),
\qquad
\mbox{if $m\gg n$} .
\eeq

Since the asymptotics for a cusp and a crack coincide, when $m \gg n$ the hypocycloidal void and the star-shaped crack tend to become mechanically equivalent, thus inducing the same stress field in the elastic plane.
This property is also highlighted in Fig. \ref{fig_crack_hypo_fullfield_comparison_curve}, where the level sets for the shear stress modulus
\beq
\tau=\sqrt{\tau_{13}^2+\tau_{23}^2} ,
\eeq
around a hypocycloidal rigid inclusion and a star-shaped
stiffener are compared and shown to coalesce at high $m$.

\end{itemize}

\subsection{Stress Reduction Factors}

In the cases of a void ($\chi=1$), loaded with $c^{(m)}_0=0$, or a
rigid inclusion ($\chi=-1$), loaded with $b^{(m)}_0=0$, the following equation holds
\beq\label{loadingconditionSRF}
b^{(m)}_0(1-\chi) -\i c^{(m)}_0(1+\chi)=0,
\eeq
and the asymptotic stress, eq. (\ref{asym_cusp}), loses the singular behaviour at the considered cusp, ($x_1=a$, $x_2=0$),
 so that the asymptotics is ruled by
the second-order term, the well-known S-stress (\cite{Moon} and \cite{radaj}). The second-order expansion
for the stress field (\ref{eq_tauconf_star_crack}) around $\zeta=1$ leads to
\begin{dmath}
\label{eq_stress_reduction}
 \tau_{13}^{(m)}-\i \tau_{23}^{(m)} = \frac{(n-1)^m a^m}{n^{m+1}}
 \sum_{j=0}^{q} \binom{m+1}{j}  \frac{(m+1-j n)^2}{(n-1)^j} \left[T+\chi \overline{T}\right] ,
\end{dmath}
so that, recalling eq (\ref{guardaunpo}),
the  stress state at the cusp point ($a,0$) of an  hypocycloidal void or a rigid
 inclusion can be obtained from eq (\ref{eq_stress_reduction}) respectively  as
\beq\label{eq_stress_reduction00}
\begin{dcases}
 \tau_{13}^{\text{\tiny\ding{71}}(m)}= \mathcal{A}(n,m) \tau_{13}^{\infty(m)}(a,0),\\
\tau_{23}^{\text{\tiny\ding{71}}(m)}=0,
\end{dcases}
\qquad
\begin{dcases}
 \tau_{13}^{\text{\tiny\ding{70}}(m)}=0,\\
\tau_{23}^{\text{\tiny\ding{70}}(m)}=\mathcal{A}(n,m) \tau_{23}^{\infty(m)}(a,0) ,
\end{dcases}
\eeq
where $\mathcal{A}(n,m)$ is the following function of the number of cusps $n$ and the loading order $m$
\begin{dmath}
\mathcal{A}(n,m) = \frac{2}{(m+1)n^{m+1}} \sum_{j=0}^{q} \binom{m+1}{j}  \frac{(m+1-j n)^2}{(n-1)^{j-m}}.
\end{dmath}

When the loading condition (\ref{loadingconditionSRF}) holds, a new parameter, the Stress Reduction Factor (SRF),
can be defined as a dimensionless measure of the stress decrease at the hypocycloidal (void or rigid) cusp point
with respect to the stress measured at the same point when the inclusion is absent,\footnote{
The concept of Stress Reduction Factor introduced here
should not be confused with an analogous terminology used in rock mechanics \cite{barton} or in seismic engineering \cite{kumar}.
}
\beq\label{SRF}
\text{SRF}(n,m):=1-\frac{\tau^{(m)}(a,0)}{\tau^{\infty(m)}(a,0)}=1-\mathcal{A}(n,m)\in[0;1),
\eeq
so that the stress state at the cusp point is described by
\beq
\tau^{(m)}(a,0)=\left[1-\text{SRF}(n,m)\right]\tau^{\infty(m)}(a,0).
\eeq

\begin{figure}[!h]
  \begin{center}
\includegraphics[width=12 cm]{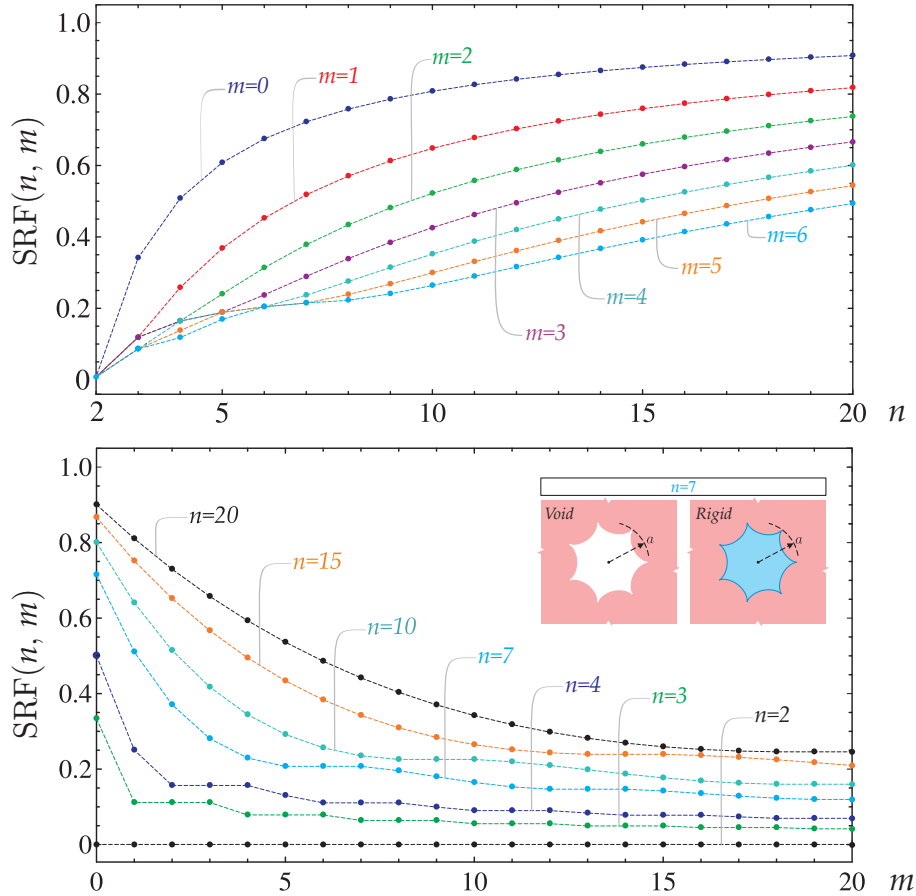}
\caption{
SRF reported as a function of the number $n$ of cusps for a hypocycloidal void or a rigid inclusion,
for different orders $m$ of the applied remote
polynomial out-of-plane shear
loading. The graphs show the properties expressed by equations (\ref{eq_srf_prop_1}) and (\ref{eq_srf_prop_2}).}
\label{fig_stress_reduction_formula}
 \end{center}
\end{figure}

Values of the SRF, eq (\ref{SRF}), are reported in Fig. \ref{fig_stress_reduction_formula}
at varying number $n$ of cusps and for different orders $m$
of the applied remote  polynomial out-of-plane shear loading.
It can be noted from the figure that two limit values are attained:
(i.) $\text{SRF} \longrightarrow 1$, when $n$ grows at fixed $m$
 (see Fig. \ref{fig_stress_reduction_formula} upper part) and (ii.)
$\text{SRF} \longrightarrow 0$, when $m$ grows at fixed $n$
(see Fig. \ref{fig_stress_reduction_formula} lower part).
These two limit values, SRF=1 and SRF=0 correspond respectively to
a stress annihilation and to an unchanged stress amount at the cusp point.
It can be observed that the latter limit value is achieved in the case of $n=2$ (linear inclusions, crack or stiffener), SRF$(n=2,m)=0$,
corresponding to the condition  of inclusion invisibility or neutrality discussed in  \cite{partII}.
It is also worth to note that the annihilation condition related to SRF=1 also occurs for a different type of inclusions, namely,
polygonal voids or rigid inclusions \cite{partII}.

The following properties for the SRF have been numerically verified (while a rigorous proof seems to be awkward) to hold
\beq
\label{eq_srf_prop_1}
\text{SRF}(n,m+1)\leq\text{SRF}(n,m)\leq\text{SRF}(n+1,m),
\eeq
\beq
\label{eq_srf_prop_2}
\text{SRF}(n,m=n)=\text{SRF}(n,m=n-1)=\text{SRF}(n,m=n-2) .
\eeq

Finally,  singularities disappear, and therefore stress reduction occurs, at all cusps of an hypocycloidal inclusion, whenever the following condition
is satisfied (for every $j \in \mathbb{N}_1$)
         \beq
                \label{eq_condizioni}
       m=\left[3-(-1)^n\right]\frac{j\,n}{4} -1.
    \eeq

\begin{figure}[!htb]
\begin{center}
\includegraphics[width=12 cm]{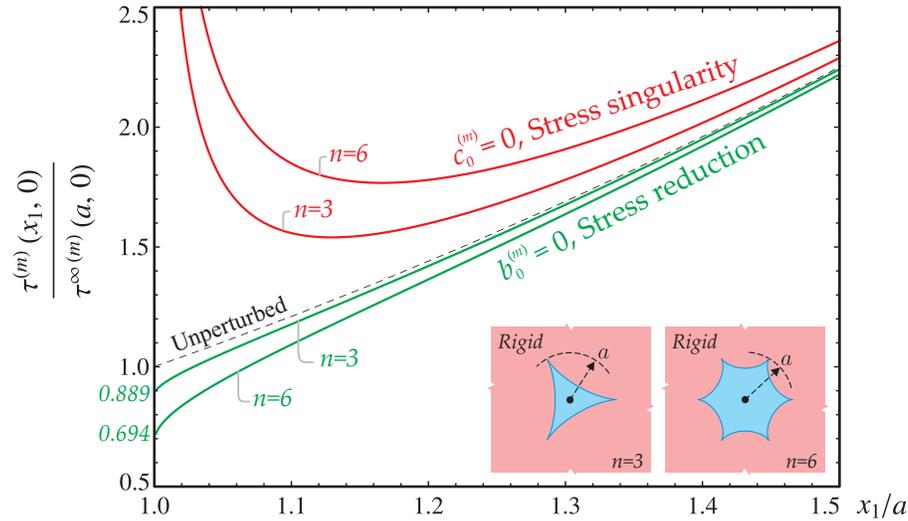}
\caption{
The shear stress modulus ahead of the cusp is reported as a function of the radial distance
from the apex of three- and six- cusped hypocycloidal inclusions for quadratic ($m = 2$)
remote out-of-plane shearing (the same case is considered in Fig. \ref{fig_stress_reduction_fullfield_and_curve_1}).
Stress singularity and stress reduction are visible at the inclusion cusp for the two considered loading conditions, $c_{0}^{(m)}=0$
and $b_{0}^{(m)}=0$, respectively.}
\label{fig_stress_reduction_fullfield_and_curve_2}
\end{center}
\end{figure}

The level sets of the dimensionless shear stress modulus $\tau^{(m)}(x_1,x_2)/\tau^{\infty(m)}(a,0)$ near rigid hypocycloidal inclusions $n=3$ (upper part) and $n=6$ (lower part)
are reported in Fig. \ref{fig_stress_reduction_fullfield_and_curve_1} for quadratic out-of-plane loading ($m=2$).
The figure highlights cases of stress singularity (left column, $c_{0}^{(m)}=0$) and stress reduction (right column, $b_{0}^{(m)}=0$).
Stress reduction vs stress singularity is also depicted in Fig. \ref{fig_stress_reduction_fullfield_and_curve_2}, through the representation of the dimensionless shear stress modulus
as a function of the distance from the cusp tip along the $x_1$-axis, for the cases  considered in Fig. \ref{fig_stress_reduction_fullfield_and_curve_1}. While the stress blows up to infinity as  a square root  singularity in the case of $c_{0}^{(m)}=0$, the stress  approaches a finite value,  smaller than that unperturbed,  in the case of $b_{0}^{(m)}=0$.

\begin{figure}[!htb]
  \begin{center}
\includegraphics[width=12 cm]{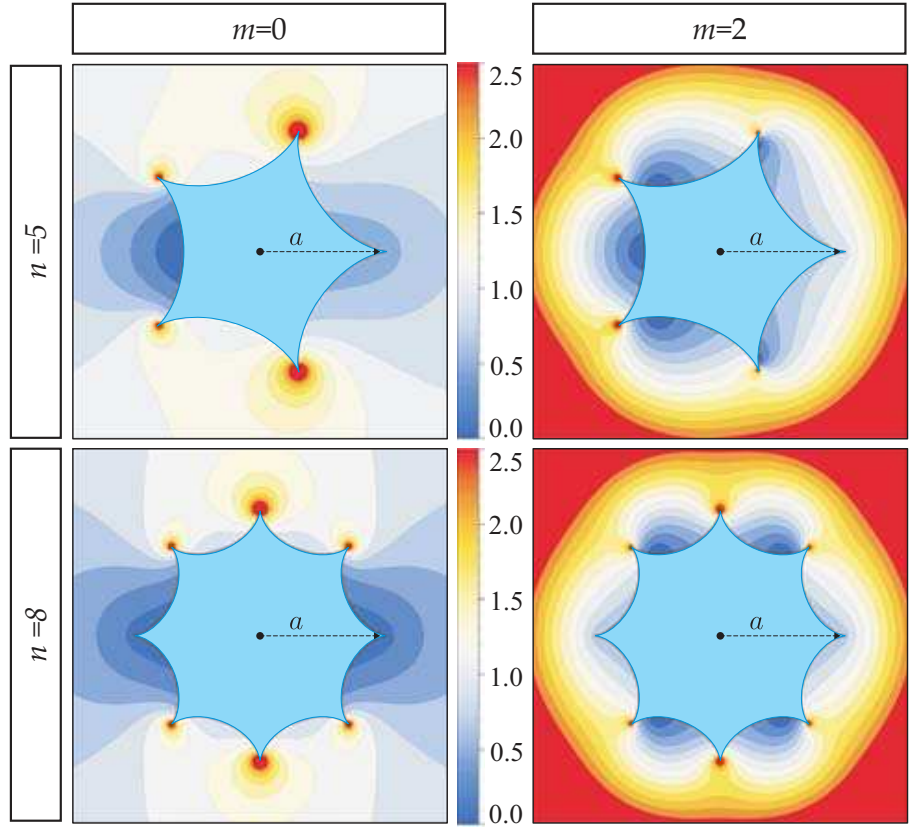}
\caption{
Level sets of dimensionless shear stress modulus $\tau^{(m)}(x_1, x_2)/\tau^{\infty(m)}(a,0)$
near $n$-cusped rigid hypocycloidal inclusions ($n=5$ and $n=8$), subject to uniform ($m=0$) and quadratic ($m=2$)
remote out-of-plane shearing ($b^{(m)}_0=0$),
show {\it partial} stress reduction, defined as the condition in which reduction occurs at some, but not all, of the inclusion vertices.
When stress reduction occurs, the
stress singularity is absent and the stress at the cusp is smaller than that which would be attained without any inclusion.
In the figures a stress singularity corresponds to the appearance of
\lq red spots' at the inclusion cusp.}
\label{fig_fullfield58}
 \end{center}
\end{figure}

Cases of \emph{partial} stress reduction occur when stress reduction is verified at some cusps, but not at the other,
namely, eq (\ref{loadingconditionSRF}) holds while eq (\ref{eq_condizioni}) does not.
Examples of such cases are reported in Fig. \ref{fig_fullfield58} for rigid hypocycloidal inclusion with five and eight cusps. Uniform ($m=0$) and quadratic
($m=2$) out-of-plane shear loadings are applied. These cases feature cusps at which the stress falls to a value smaller than that would be attained at the  same point in the absence of  the
inclusion, while stress singularities are still present at the other cusps.

\section{Conclusions}

Voids and rigid inclusions with hypocycloidal
shapes embedded in
an infinite elastic plane subject to remote polynomial antiplane loading were analyzed.
The solutions to these problems were solved in terms of full-field representation and stress intensity factors.
It has been shown that in certain situations cusps
can act to reduce stress instead of increasing it, as might be expected.
This effect has been quantified by introducing the notion of Stress Reduction Factor.
Moreover, it has been shown that when the order of the applied antiplane shear load is increased, the hypocycloidal-shaped inclusion generates a stress field that tends to coincide with the stress field generated by a star-shaped crack or stiffener.
The obtained results can be used as a guide for the design of composites with superior mechanical characteristics.

\section*{Acknowledgments}
The authors gratefully acknowledge financial support from the ERC Advanced Grant \lq Instabilities and nonlocal multiscale modelling of materials'
ERC-2013-ADG-340561-INSTABILITIES.


\begin{thebibliography}{99}


\bibitem{ammari}
Ammari, H., Kang, H., Lee, H., Lim, J., 2013. Boundary perturbations due to the presence of small linear cracks in an elastic body.
\emph{J. Elasticity}  113,  1, 75--91.


\bibitem{ammari2}
Ammari, H., Ciraolo, G., Kang, H., Lee, H., Yun, K., 2013.
 Spectral analysis of the Neumann-Poincaré operator and characterization of the stress concentration in anti-plane elasticity.
\emph{ Arch. Ration. Mech. Anal.}  208, 1, 275--304.

\bibitem{bacca} Bacca, M., Bigoni, D., Dal Corso, F., Veber, D.,  2013.
Mindlin second-gradient elastic properties from dilute two-phase Cauchy-elastic composites Part II: Higher-order constitutive properties and application cases. \IJSS 50, 4020-4029.

\bibitem{barbieri} Barbieri, E., Pugno, N.M., 2015.
A computational model for large deformations of composites with a 2D
soft matrix and 1D anticracks. \IJSS 77, 1-14.

\bibitem{barton} Barton, N., Lien, R. Lunde, J., 1974.
Engineering Classification of Rock Masses for the Design of Tunnel Support.
Rock Mechanics 6, 189-236

\bibitem{chen} Chen, Y.Z., 2013. Evaluation of T-stress for a hypocycloid hole in an infinite plate. Multidiscipline Model. Materials Struct. 9, 450-461.

\bibitem{chen2} Chen, Y.Z., 2014. Eigenfunction expansion variational method for the solution of a cusp crack problem in a finite plate. Acta Mech.
168, 157-166.


\bibitem{craciun}
Craciun, E.M., Soós, E., 1998. Interaction of two unequal cracks in a prestressed fiber reinforced composite.
\IJF, 94, 137-159.





\bibitem{hasebe_multiple_crack} Chen, Y.Z., Hasebe, N., and Lee, K.Y., 2003. Multiple crack problems in elasticity. WIT press.

\bibitem{chu} Chu, C.M., 1994. The stress field singularity near a cusp. \EFM 47, 361-365.

\bibitem{stiffener0} Dal Corso, F., Bigoni, D., Gei, M., (2008). The stress concentration near a rigid line inclusion in a prestressed, elastic material.
Part I. Full field solution and asymptotics. J. Mech. Phys. Solids, 56, 815-838.

\bibitem{partI} Dal Corso, F., Shahzad, S., Bigoni D., 2016.  Isotoxal  star-shaped  polygonal  voids  and  rigid  inclusions  in  nonuniform
antiplane  shear  fields.  Part  I:  Formulation  and  full-field
solution.  \IJSS 85-86, 67-75.

\bibitem{partII} Dal Corso, F., Shahzad, S., Bigoni D., 2016.  Isotoxal  star-shaped  polygonal  voids  and  rigid  inclusions  in  nonuniform
antiplane  shear  fields.  Part  II:  Singularities,  annihilation  and  invisibility.  \IJSS 85-86, 76-88.

\bibitem{gao} Gao H., 1995. Mass-conserved morphological evolution of hypocycloid cavities: a model of diffusive crack initiation with no associated energy barrier. Proc. R. Soc. Lond. A 448, 465-483.

\bibitem{gdoutos} Gdotus, E.E., 2003. Problem 68: Failure of a plate with a hypocycloidal inclusion, In Problems of Fracture Mechanics and Fatigue, E. Gdoutos, C.A. Radopoulos and J.R. Yates Editors, Springer.

\bibitem{panos} Gourgiotis, P.A., Piccolroaz, A., 2014. Steady-state propagation of a mode II crack in couple stress elasticity.
Int. J. Frac. 188 (2), 119-145.

\bibitem{hasebe} Hasebe, N., Nemat-Nasser, S., Keer, L.M., 1984. Stress analysis of a kinked crack initiating from a rigid line inclusion.
Part II: Direction of propagation. Mech. Mat., 3 (2), 147-156

\bibitem{dascalu} Homentcovschi, D., Dascalu, C., 2000. Uniform asymptotic solutions for lamellar inhomogeneities in plane elasticity. J. Mech. Phys. Solids, 48, 153-173

\bibitem{japan} Kohno, Y., Ishikawa, H., 1995. Singularities and stress intensities at the corner point of a polygonal hole and rigid polygonal inclusion under antiplane shear. \IJES 33, 1547-1560.

\bibitem{ivanov} Ivanov, V.I., Trubetskov, M.K., 1994. Handbook of Conformal Mapping with Computer-Aided Visualization. CRC Press.

\bibitem{inclusioni} Misseroni, D., Dal Corso, F., Shahzad, S., Bigoni, D., 2014. Stress concentration near stiff inclusions:
Validation of rigid inclusion model and boundary layers by means of photoelasticity. \EFM 121-122, 87-97.

\bibitem{kumar} Kumar, J. 2006.
Stress Reduction Coefficient and Amplification Factor
for Seismic Response of Ground.
Int. J. Geomech. 6, 141-146.

\bibitem{misra} Misra, S., Mandal, N., 2007. Localization of plastic zones in rocks around rigid inclusions: insights from experimental and theoretical models. J.  Geophys.  Res. 112, B09 206.

\bibitem{Moon} Moon, H.J., Earmme, Y.Y., 1998. Calculation of elastic T-stresses near interface crack tip under in-plane and anti-plane loading. \IJF 91, 179-195.

\bibitem{movmov} Movchan, A.B., Movchan, N.V., 1995. Mathematical Modelling of Solids with Nonregular Boundaries. CRC Press.

\bibitem{movchan} Movchan, A.B., Movchan, N.V., Poulton, C.G., 2002. Asymptotic Models of Fields in Dilute and Densely Packed Composites. Imperial College Press.

\bibitem{long}  Nik Long, N.M.A., Yaghobifar, M., 2011. General analytical solution for stress intensity factor of a hypocycloid hole with many cusps in an infinite plane. Philosophocal Magazine Letters 91:4, 256-263.

\bibitem{noselli} Noselli, G., Dal Corso, F. and Bigoni, D., 2010. The stress intensity near a stiffener disclosed by photoelasticity. \IJF 166, 91-103.

\bibitem{ozturk} Ozturk, T., Poole, W.J., Embury, J.D., 1991. The deformation of Cu-W laminates. Mater. Sci. Eng. A 148, 175-178.


\bibitem{trush} Panasyuk, V.V., Berezhnitskii, L.T. and Trush, I.I., 1972. Stress distribution about defects such as rigid shar-angled inclusions. Physicomechanical institute of the academy of science of Ukrainian SSR, L'vov. Translated from Problemy Prochnosti 7, 3-9.

\bibitem{pana2} Panasyuk, V.V., Buina, E.V., 1966. Threshold equilibrium of a plate weakened by a polygonal hole. Fiziko-Khimicheskaya Mekhanika Materialov 2, 15-20.

\bibitem{pana3} Panasyuk, V.V., Buina, E.V., 1967. Critical stress diagrams for brittle materials with defects of the cusped void/crack type. Fiziko-Khimicheskaya Mekhanika Materialov, 3, 584-591.

\bibitem{roaz} Piccolroaz, A., Mishuris, G., Movchan, A.B., 2010.
Perturbation of mode III interfacial cracks. Int. J. Frac. 166, 41-51.

\bibitem{radaj} Radaj, D., 2013. State-of-the-art review on extended stress intensity factor concepts. Fatigue Fract. Eng. Mat. Str. 37, 1-28.

\bibitem{radi}  Radi, E., 2007. Effects of characteristic material lengths on mode III crack propagation in couple stress elastic-plastic materials.
Int. J. Plast. 23 (8), 1439-1456.

\bibitem{schiavone}  Schiavone, P., 2003. Neutrality of the elliptic inhomogeneity in the case of non-uniform loading. \IJES 8, 161-169.

\bibitem{sih} Sih, G.C., 1965. Stress Distribution Near Internal Crack Tips for Longitudinal Shear Problems. \JAM 32, 51-58.

\bibitem{vasudevan} Vasudevan, M., Schiavone, P., 2006. New results concerning the identification of neutral inhomogeneities in plane elasticity. Arch. Mechanics 58, 45-58.

\bibitem{vigder1} Vigdergauz, S., 2007. Shape-optimization of a rigid inclusion in a shear-loaded elastic plane. \JoMMS 2, 275-291.

\bibitem{vigder2} Vigdergauz, S., 2008. Energy-minimizing inclusion in an elastic plate under remote shear. \JoMMS 3, 63-83.

\bibitem{wang} Wang, X., Schiavone, P. 2014. Finite deformation of harmonic solids with cusp cracks. IMA J. Appl. Math. 79, 790-803.


\end{thebibliography}
\end{document}